# Photodesorption efficiency of OH radical on the ice surface in the wavelength range from ultraviolet to visible

Ni-En Sie [a], Masashi Tsuge [a], Yoichi Nakai [b], Naoki Watanabe [a]

[a] *Hokkaido University, Sapporo, Hokkaido, 060-0819, Japan*
[b] *RIKEN Nishina Center for Accelerator-based Science, Wako, Saitama, 351-0198, Japan*

Abstract

The photodesorption efficiencies of the hydroxyl (OH) radical from the water ice surface were measured in the range of 310–700 nm for the first time. Although isolated $H_2O$ molecules and OH radicals do not absorb visible photons, the photodesorption of OH interacting with the ice surface was found to occur by a one-photon process in the entire visible range. The photodesorption efficiency strongly depended on the wavelength, and the analyses of photodesorption cross sections indicated that only strongly bound OH radicals were desorbed at longer wavelengths, whereas both weakly and strongly bound OH radicals were desorbed at shorter wavelengths.

## 1. Introduction

OH radicals play an important role in atmospheric and interstellar chemistry because of their high reactivity and large abundance owing to rich formation pathways such as the photolysis of water molecules on ice dust particles. For instance, in interstellar environments where the temperature can reach 10 K, the OH radicals can remain on the surface of interstellar ice dust and contribute to the formation of a variety of interstellar molecules, such as $CO_2$ from CO + OH reaction, $CH_3OH$, and $H_2O_2$ [1-3]. To better understand OH chemistry under various conditions, the elementary processes of OH on ice, namely, surface diffusion and desorption, need to be determined [4, 5]. Unlike a single-crystal surface, the surfaces of polycrystalline and amorphous ices have many kinds of OH radical adsorption sites, causing difficulty in both experimental and computational investigations. Although the thermal desorption of OH has not been observed experimentally, the binding energies of OH with various adsorption sites on ice surfaces were determined by quantum chemical calculations to be approximately 0.37 eV on average [4]. The photodesorption of OH upon the photolysis of $H_2O$ in the vacuum ultraviolet region has been extensively studied *via* both experiments and molecular dynamics (MD) simulations [6-8]. Although the OH photodesorption from ice at 90 K was experimentally observed upon irradiation at 157 nm, the MD calculation indicated that more than 90% of OH radicals do not desorb upon photodissociation of $H_2O$; this occurred because most of the excess energy was used by the H-atom desorption [7, 9]. To better understand the chemistry of OH on ice, selective monitoring of OH radicals on the ice surface is desirable. However, conventional methods that are often used for solids, such as Raman, infrared, and electron spin resonance spectroscopies, are not applicable because of their inability to selectively analyze species on the surface. Furthermore, due to the high reactivity, it is difficult to prepare a significant surface number density of OH on ice.

Therefore, highly sensitive methods are needed. Recently, we were successful in detecting radicals on ice with a combination of photostimulated desorption (PSD) and resonance-enhanced multiphoton ionization (REMPI) techniques, known as the PSD-REMPI method [10-13]. Using this method, the OH radicals on ice prepared by ultraviolet (UV) photolysis of $H_2O$ ice could be detected, and the behaviors were monitored at temperatures between 54 and 80 K [5]. The OH surface number density in a steady state during UV exposure gradually decreased at temperatures above 60 K. Analyzing the temperature dependence of OH intensities with the Arrhenius-type equation, the decrease could be explained by the recombination of two OH radicals; this process was rate-limited by the thermal diffusion of OH. The activation energy for surface diffusion was experimentally determined to be 0.14 ± 0.01 eV [5].

The development of the PSD-REMPI method for detecting OH radicals on ice led to additional findings that the photodesorption of OH from ice was stimulated by the PSD laser radiation at 532 nm, where both the isolated OH radical and $H_2O$ molecule are transparent [14-17]. In Ref. [4], OH radicals were found to photodesorb *via* a one-photon process at 532 nm. For the isolated OH radical, excitation to the first doublet excited state, i.e., the $^2\Sigma^+ \leftarrow {}^2\Pi$ valence transition (*A-X* transition), occurs at 308 nm [18, 19]. However, quantum chemical calculations have shown that the electronic *A-X* transition can occur for OH radicals adsorbed on ice surfaces in the visible range of approximately 600 nm [4]. The photon wavelength for the transition was found to depend on the OH adsorption sites on the surface of the water ice. In particular, when OH has three hydrogen bonds with the neighboring three $H_2O$ molecules on ice, the transition can occur at a relatively longer wavelength near 532 nm. Although the details of the desorption mechanism are still unclear, the excited-state structure near the potential minimum can potentially enter a dissociative channel through a conical intersection. A theoretical approach for clarifying the pathways for desorption from ice is certainly challenging. The correlation between the OH desorption efficiency and photon wavelength may provide information on either the absorption cross sections of *A-X* transitions or the population of adsorption sites. In the present study, we measured the photodesorption efficiency of OH radicals on a water ice surface, which shows a drastic wavelength dependence from the UV to the visible region. In addition, the photodesorption cross sections are measured at different wavelengths, and our results can provide further insight into future theoretical studies.

## 2. Experiments

The experiments were conducted in an ultrahigh vacuum chamber with a base pressure of $3 \times 10^{-7}$ Pa; this was equipped with a helium cryostat to cool the substrate to the lowest temperature of 10 K. A sapphire crystal with a diameter of 6 mm was installed with high-temperature conductive silver paste onto the center of the aluminum substrate with a diameter of 40 mm. The water ice in so-called compact amorphous solid water (ASW) was formed on a sapphire substrate by background vapor deposition at 100 K, and the thickness of ~270 monolayers was estimated from the increase in pressure to $2 \times 10^{-5}$ Pa with the $H_2O$ vapor and a deposition duration of 30 minutes. The OH radicals were produced by the exposure of water ice at 50 K to UV photons at 45 degrees incident from a conventional deuterium lamp; this generated photons in the range of 115–400 nm with a flux of approximately $10^{13}$ photons $cm^{-2}$ $s^{-1}$, which was measured by a photodiode (AXUV-100G, IRD Inc.). The irradiation duration was changed for different sets of experiments. UV photons above ~7.3 eV (170 nm) dissociates $H_2O$ mainly into H + OH with minor channels such as $H_2$ + O and 2H + O [6, 20, 21]. However, with the exception of OH, these photofragments cannot stay on the ice surface at 50 K; this temperature is far above the desorption temperatures of H, H$_2$, O, and O$_2$. However, considering the activation energy of 0.14 eV for OH surface diffusion, a significant amount of OH radicals could remain intact on ice at 50 K during the present experimental duration [5].

Figure 1 shows an overview of the experimental procedure. Briefly, OH radicals on ASW were photodesorbed by irradiation with a Nd$^{3+}$:YAG pumped optical parametric oscillator laser (hereafter, PSD laser). The spectral width of the wavelengths in the range of 310–700 nm was 4 cm$^{-1}$, and the laser powers were approximately set under 0.2 mJ pulse$^{-1}$ with a beam spot of ~3 mm$^2$ on the sapphire disk. The photodesorbed OH was subsequently photoionized at typically 1 mm above the ice surface by the (2+1) REMPI process *via* the $D^2\Sigma^- \leftarrow X^2\Pi$ transition of the OH radical [22]. The photons in the range 243.9–244.9 nm for the REMPI process were provided by a Nd$^{3+}$:YAG pumped pulsed dye laser (PrecisionScan, Sirah). The ionized OH was eventually detected by a time-of-flight mass spectrometer (TOF-MS). The lasers were operated at 10 Hz during the data acquisition. The photodesorption of OH was investigated in two different modes: during UV irradiation and after terminating UV irradiation, as illustrated in Figs. 1(d) and (e), respectively.

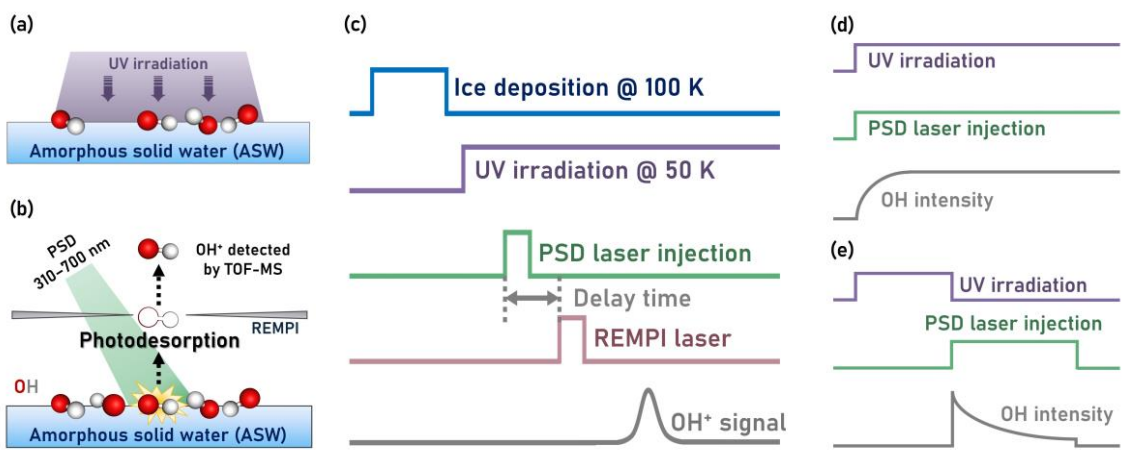

**Fig. 1.** (a) OH production from the UV photolysis of water ice. (b) Process of OH photodesorption and the subsequent REMPI process. (c) Timing charts of OH signal detection. (d) and (e) Timing charts of the OH intensity during UV irradiation and after terminating UV irradiation.

## 3. Results and Discussion

Figure 2 shows the experimentally obtained (2+1) REMPI spectra during UV irradiation at four wavelengths of the PSD laser together with the simulated spectra of OH obtained by the PGOPHER program [23]. Vibrationally excited OH (v″ = 1) was not observable, and the profiles were quite similar at each wavelength, although the signal-to-noise ratios differed. Regardless of the wavelength, the experimentally obtained spectra could be effectively reproduced by calculations with rotational temperatures in the range of 120–200 K; this result is consistent with previous experiments at 532 nm [4]. Here, a simulation with a 140 K configuration was selected for comparison with experimental data collected at 50 K. These rotational temperatures (120–200 K) were not consistent with the ice temperature at 50 K, and a discrepancy was also observed in a previous experiment where the OH fragments from the photodissociation of H$_2$O nm on ice at 90

K were detected [7]. Although the mechanism for determining these specific rotational temperatures is unclear, it should certainly be affected by desorption dynamics. Hereafter, the highest REMPI intensity at 244.162 nm for the $R_1$ ($J = 1$) branch [22] was used to monitor the OH photodesorption intensity.

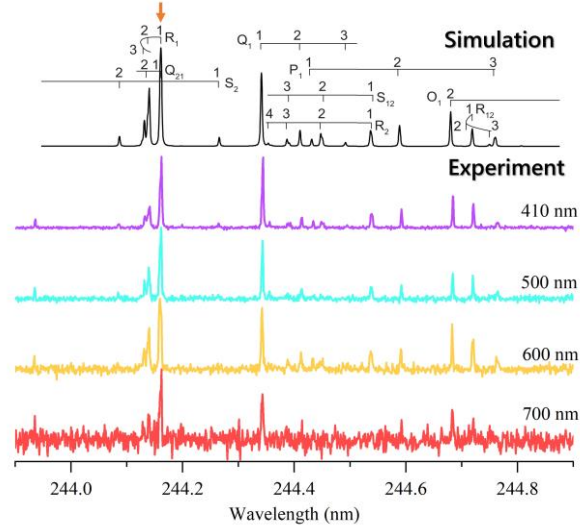

**Fig. 2.** Experimental (2+1) REMPI spectra of the OH radicals desorbed from ASW by a 410–700 nm PSD laser compared with the simulated spectrum of the $D^2\Sigma^-$ ($v' = 0$) ← $X^2\Pi$ ($v'' = 0$) transition at a rotational temperature of 140 K. An arrow on the top spectrum shows the REMPI wavelength 244.162 nm for monitoring the OH intensity.

The OH radicals were generated by the dissociation of $H_2O$ by UV lamp irradiation and accumulated on the ASW surface at 50 K. Continuous UV irradiation for approximately 20 minutes was found to produce a steady state in the number density of OH radicals on the ice surface. The steady-state number density of OH radicals needed to be determined by the dynamic equilibrium between OH formation by $H_2O$ dissociation and OH loss by UV lamp and PSD laser induced photodesorption and/or recombination reaction. According to Ref. [5], the formation of $H_2O_2$ by the diffusive OH–OH recombination predominantly occurs at temperatures higher than 60 K with an activation energy of 0.14 eV. We also confirmed that the OH intensity remained at the same level when the ice was left in the dark without irradiation by UV lamp or PSD laser at 50 K. Specifically, OH attenuation by thermally activated recombination of OH was not observable in this study. The OH number density, [OH], on the ice surface under continuous UV irradiation can be approximately expressed by setting the following equation to zero:

$$\frac{d[OH]}{dt} = \alpha f_{UV}\sigma_{diss}[H_2O] - f_{UV}\sigma_{loss}[OH] - f_{PSD}\sigma_{des}(\lambda)[OH], \tag{1}$$

where $\alpha$ is the fraction of OH remaining on the surface at photodissociation, $f$ is the photon flux of the UV lamp or PSD laser, $\sigma_{diss}$ is the UV dissociation cross section of water ice, $\sigma_{loss}$ represents the total loss cross section of OH by both photodesorption and photodissociation (threshold 4.39 eV) induced by the UV lamp [16], and $\sigma_{des}(\lambda)$ is the photodesorption cross section by PSD laser irradiation at wavelength $\lambda$. For reference, the UV-absorption cross section of water ice at 120–160 nm is approximately $5 \times 10^{-18}$ cm$^2$ [24]. In Eq. 1, only the last term of the right side depends on the wavelength of the PSD laser.

## 3.1. Photodesorption efficiency

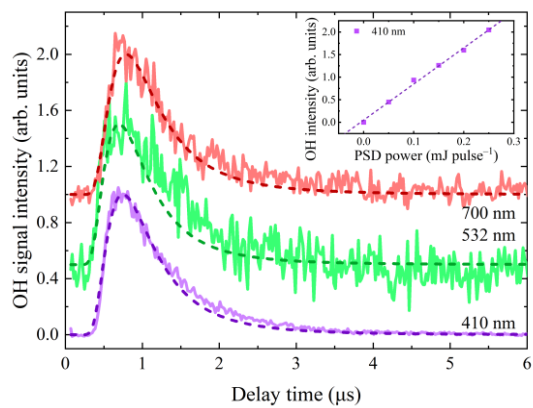

**Fig. 3.** Translational energy distribution for photodesorbed OH radicals at PSD wavelengths of 410, 532, and 700 nm, where the delay time corresponds to the flight time of OH radicals between the ice surface and focal point of the REMPI laser. The dashed lines show the results of Maxwell–Boltzmann distribution fittings. The inset shows the PSD power dependence of OH intensities at 410 nm.

The photodesorption efficiency was measured at the steady state under continuous UV irradiation, as shown in the timing chart in Fig. 1(d). By changing the delay time between the PSD and the REMPI lasers, the OH intensities were measured as a function of delay time, as shown in Fig. 3. The delay time corresponded to the flight time of OH radicals for the distance between the ice surface and the focal point of the REMPI laser. The measured profiles reflect the translational energy distributions of OH radicals desorbed from the ice surface, and these were well fitted with single Maxwell–Boltzmann distribution functions with temperatures from 1800 K to 2650 K following the manner described previously [7, 25, 26]. These translational temperatures were similar to the values experimentally obtained in Ref. [4]. The reason for the observed variation at different wavelengths is unclear, and extensive excited state calculations are required to get further insights.

The area intensities obtained by integration of the plots in Fig. 3 are proportional to the number of desorbed OH radicals per PSD laser pulse. We confirmed that the area intensities are linearly correlated with the PSD laser power for all the wavelengths of the PSD laser (as an example, the case for 410 nm is shown as an inset of Fig. 3); these results indicate that photodesorption was induced by a one-photon chemical process rather than phonon propagation-induced desorption along with a power-law dependence on the PSD laser power [4]. The area intensities divided by the photon flux were calculated as the photodesorption efficiencies, as shown in Fig. 4(a).

We found that the photodesorption efficiency of OH adsorbed on the water ice surface strongly depends on the wavelength, and the maximum desorption efficiency falls within 310–430 nm. In the visible region, the photodesorption efficiency at 410 nm is the largest, decreasing by approximately one order of magnitude at 532 nm, and a small bump is observed at approximately 600–650 nm. At 700 nm, the photodesorption efficiency is approximately 40 times lower than that for the 410 nm configuration. Figure 4(b) shows the calculated *A-X* transition wavelengths for various adsorption sites as a function of the binding energy between OH and the ice surface. We performed additional calculations to increase the sampling of binding sites by using the same method that has been employed in Ref. [4]. As shown in Fig. 4(b), the transitions are predicted to predominantly appear within the wavelength range from 300 to 410 nm, which is likely the reason that the photodesorption efficiencies in the range of 310–430 nm are greater than those in the longer wavelength range of the visible region. However, information on the excitation wavelength is insufficient for quantitatively explaining the trend of the wavelength-dependent efficiency. One should note that the sampling from the quantum chemical calculation is not complete, and we do not have a clear explanation to describe the small bump at near 600 nm on the photodesorption efficiency. The photodesorption efficiency at each wavelength depends on three factors. First, the population of adsorption sites where the *A-X* transition can occur. The second and third ones that control the photodesorption efficiency are the photoabsorption cross section and the quantum yield for photodesorption. These factors cannot be separated experimentally but can be combined into photodesorption cross sections. Therefore, we measured the photodesorption cross section to gain some insights into the wavelength-dependence characteristics.

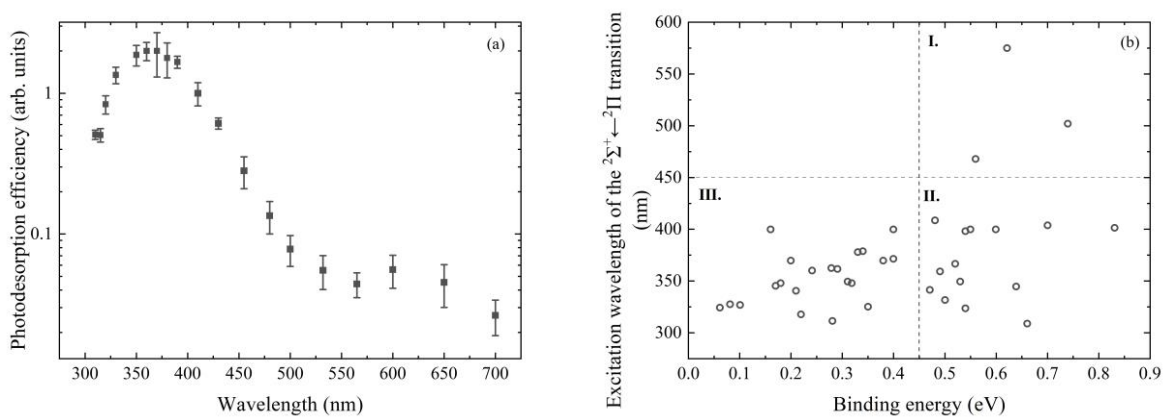

**Fig. 4.** (a) Relative photodesorption efficiency as a function of the PSD laser wavelength, where photodesorbed OH radicals were ionized by the REMPI laser at 244.162 nm. (b) Excitation wavelength of the $^2\Sigma^+ \leftarrow {}^2\Pi$ transition plotted against the binding energy from quantum chemical calculations, employing the ice models in a previous study [4] and using ωB97X-D/6-31G(d) level of theory. The numbers I, II, and III represent the areas divided by binding energy and excitation wavelength (see text).

*3.2. Photodesorption cross section*

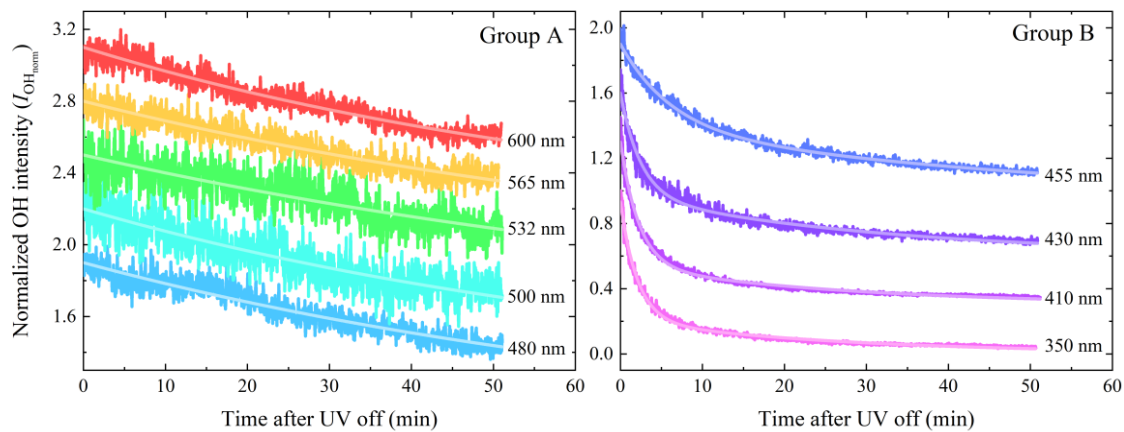

**Fig. 5.** Normalized OH intensity ($I_{OH_{norm}}$) for ices at 50 K as a function of time after the termination of UV lamp irradiation for different wavelengths. Groups A and B showed distinct decay trends (see text). Note that the data were vertically offset for clarity.

In the series of experiments, the ice was exposed to a UV lamp for 20 min to accumulate OH radicals on the surface at 50 K; at this temperature, the loss of OH radicals due to recombination reactions induced by their thermal diffusion is negligible [5]. Subsequently, the UV radiation was turned off, the sample was irradiated with a PSD laser, and the temporal variations in the OH intensity were measured (see the scheme shown in Fig. 1e). The delay time between the PSD and REMPI lasers was set at around 0.7 μs which corresponds to timing providing the maximum intensity of each OH translational energy distribution (see Fig. 3). The peak positions of the translational energy distribution did not change with time after UV termination. During continuous injection of the PSD laser, the OH number density decreased due to the photodesorption induced by the PSD laser. Note that PSD laser radiation did not induce the photodissociation of $H_2O$ to produce OH. The detected OH intensity is proportional to the surface number density and photodesorption cross section at each wavelength of the PSD laser, and the time variation of OH intensities was traced for 50 minutes for eleven PSD wavelengths.

Figure 5 shows the attenuation of normalized OH intensity, $I_{OH_{norm}}$, measured after UV lamp irradiation was terminated. The profiles of OH attenuation can be categorized into two different groups from decay trends: these are referred to as Group A (longer than 480 nm) and Group B (shorter than 455 nm). In Group A, the decay trends can be well represented by a single exponential function. In contrast, the OH intensities for Group B cannot be effectively fitted with a single exponential component only, and at least two exponential components are needed to better fit the decay curves. We can interpret this as follows: for photons at shorter wavelengths, two different kinds of adsorption sites likely exist leading to desorption, whereas, for longer wavelengths, the adsorption sites responsible for desorption are represented by a single kind. With this concept, Fig. 4(b) shows that for wavelengths longer than ~450 nm, the excitations occur only when OH is adsorbed on a high binding energy site (≥0.45 eV) on the ice surface (area I), while the excitation at shorter wavelengths below 450 nm covers both high and lower binding energy sites (areas II and III). The cross sections of photodesorption may differ between OHs on strong and weak binding sites. Quantum chemical calculations by Ref. [27] demonstrated eight optimized structures of the OH–($H_2O$)$_n$ complexes with individual binding energies in the range of 0.20–0.67 eV, indicating that numerous kinds of adsorption sites were present on the water ice surface; however, it is challenging for both experiments and quantum chemical calculations to clarify the exact population. Here, we simply divide them into two groups by a binding energy of 0.45 eV.

Similar to Eq. 1, the number densities for two kinds of adsorbed OH without UV lamp irradiation can be written by the following equations:

$$\frac{d[OH]_1}{dt} = -f_{PSD}\sigma_{des-1}(\lambda)[OH]_1, \qquad (2)$$

and

$$\frac{d[OH]_2}{dt} = -f_{PSD}\sigma_{des-2}(\lambda)[OH]_2. \qquad (3)$$

The $[OH]_1$ and $[OH]_2$ represents the OH radicals at strong and weak binding sites, respectively, while the summation of them equals the total OH surface number density [OH]. The $\sigma_{des-1}$ and $\sigma_{des-2}$ are the corresponding photodesorption cross sections for strong and weak binding sites, respectively. The OH intensity in the experiment, $I_{OH}$, is determined by the surface number density ([OH]), the photodesorption cross section, and the REMPI yield. Because the REMPI yield can be considered constant, by multiplying the number density with the photodesorption cross section, the OH number densities at strong and weak binding sites, $[OH]_1$ and $[OH]_2$, are related to $I_{OH}$ as $I_{OH} \propto \sigma_{des-1}[OH]_1 + \sigma_{des-2}[OH]_2$. Thus, by solving Eqs. 2 and 3, the normalized OH intensity, $I_{OH_{norm}}$, is represented by

$$I_{OH_{norm}}(t) = \sigma_{des-1} P_1 e^{-f_{PSD}\sigma_{des-1}(\lambda)t} + \sigma_{des-2} P_2 e^{-f_{PSD}\sigma_{des-2}(\lambda)t}, \qquad (4)$$

where $P_1$ and $P_2$ represent the relative populations of strong and weak binding adsorption sites at $t = 0$, respectively. In the case of Group A, since the attenuation of OH intensity as a function of time without UV irradiation represents a single exponential decay (Fig. 5), we omitted the second term of Eq. 4 for a single exponential fitting.

By fitting the OH intensity as a function of time with Eq. 4 for both groups, the population of adsorption sites ($P_{1,2}$) and photodesorption cross sections for strong and weak binding ($\sigma_{des-1,2}$) could be obtained. For longer wavelengths (Group A), the cross section $\sigma_{des-1}$ was determined to be $\sim(0.2-0.5) \times 10^{-22}$ cm$^2$, while for shorter wavelengths, two cross sections were determined to be $\sim(0.6-1.8) \times 10^{-22}$ cm$^2$ and $\sim(0.60-1.98) \times 10^{-21}$ cm$^2$. Mathematically, the two exponential terms in Eq. 4 for Group B are indistinguishable whether they are due to strong or weak binding sites. However, by comparing with the values of $\sigma_{des-1}$ for Group A, the term for the strong binding sites can be assumed to share a similar value to that at longer wavelengths, as shown in Fig. 6(a). The higher cross section values are attributed to photodesorption of the OH radical from the weak binding sites. The values of $P_1$ and $P_2$ were determined to have comparable values of 0.39–0.76 and 0.24–0.61, respectively. These populations are consistent with the predicted correlation between the *A-X* excitation wavelength and the binding energy of OH–(H$_2$O)$_n$ shown in Fig. 4(b) that the calculated number of excitation points below and above 0.45 eV are quite similar. The figure is divided into four areas by a horizontal line at a wavelength of ~450 nm and a vertical line at a binding energy of 0.45 eV; here, the longer wavelength with strong binding energy is area I, and the shorter wavelengths with strong and weak binding energies are areas II and III, respectively. The $\sigma_{des-1}$ determined for longer wavelengths would be originated mainly from the sites in area I, and the $\sigma_{des-1}$ and $\sigma_{des-2}$ for shorter wavelengths originated from areas II and III, respectively. The comparable values of $P_1$ and $P_2$ might be explained by the similarity in the number of sampled points in areas II and III.

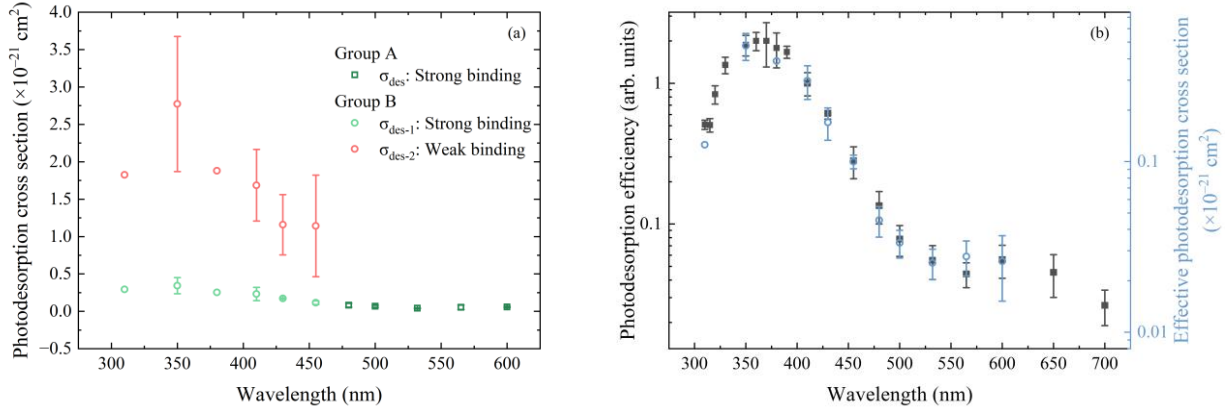

**Fig. 6.** (a) Photodesorption cross sections of OH from the strong and weak binding sites derived from Eq. 4. Dark green squares indicate long wavelengths from strong binding sites; green circles indicate short wavelengths from strong binding sites; red triangles indicate short wavelengths from weak binding sites. (b) Relative photodesorption efficiency from Fig. 4(a) and the effective photodesorption cross section as a function of wavelength.

To compare the results of photodesorption cross section measurements with those of photodesorption efficiency measurements, we introduce the effective photodesorption cross section $\sigma_{\text{eff}}(\lambda)$ as weighted averages of the cross sections for the population of two binding sites given by

$$\sigma_{\text{eff}} = \frac{P_1}{P_1+P_2} \times \sigma_{\text{des-1}}(\lambda) + \frac{P_2}{P_1+P_2} \times \sigma_{\text{des-2}}(\lambda). \tag{5}$$

It should be noted that for $\lambda \geq 480$ nm (Group A), the second term of Eq. 4 is negligible because the OH photodesorption occurred only from strong binding sites (single exponential component). Therefore, the effective cross section for $\lambda \geq 480$ nm is equivalent to $\sigma_{\text{des-1}}(\lambda)$. The effective photodesorption cross sections are shown by hollow circles in Fig. 6(b), which has a trend identical to that of the photodesorption efficiency as a function of wavelength. The determined photodesorption cross sections are in the order of $\times 10^{-21} - 10^{-22}$ cm$^2$, which is much smaller compared to the absorption cross section $8.78 \times 10^{-17}$ cm$^2$ of OH at 308 nm [18]; thus the quantum yield is very low for the photodesorption of OH radicals in this wavelength region. The comparable $P_1$ and $P_2$ values (0.39–0.76 and 0.24–0.61, respectively) suggest that the populations of OHs on strong and weak binding sites are similar to each other. Consequently, this result confirms that the variation of photodesorption efficiency at different wavelengths is consistent with the multiplication product of photoabsorption cross section and quantum yield. Unfortunately, the present experiments cannot clarify the reason for observed features in the wavelength-dependent photodesorption efficiency curve, such as a peak at ~370 nm and a small bump at ~600 nm (Fig. 6b). More extensive quantum chemical calculations are necessary for understanding the physical meanings of these characteristics.

## 4. Summary

Unlike UV photons and energetic X-rays, visible light was not expected to trigger the photodesorption of OH from the surface of H$_2$O ice since water and OH radicals do not absorb visible light. Through the PSD-REMPI method, we successfully detected photodesorbed OH radicals from the ice surface induced by visible light, which shows the photon absorption of the OH radicals on the water ice surface in the visible region. The photodesorption efficiency is highly wavelength correlated, with a maximum at ~370 nm. In the UV region below 400 nm, the desorption efficiency is remarkably greater than that in the visible light region; this is based on the quantum chemical results showing that most of the excitation occurs within 300–400 nm. In the visible region, the efficiency decreases as the wavelength increases. Furthermore, the measurement of the photodesorption cross section led to an improved understanding of the relationship between photon absorption and OH adsorption sites on the ice surface; both of these control photodesorption processes. OH radicals at different adsorption sites on the water ice surface were photodesorbed by impinging photons of different wavelengths. We found that short wavelength photons (≤455 nm) could be absorbed by the OH radical, likely at weak binding adsorption sites and at strong binding sites. Conversely, longer wavelengths would be only absorbed by strong binding sites (≥0.45 eV), resulting in photodesorption. The fitting results showed a comparable population of strong and weak binding sites, and the effective photodesorption cross section effectively reproduced the wavelength-dependent desorption efficiency. Besides the population of adsorption sites, the photoabsorption cross section and quantum yield also influence the photodesorption efficiency. By the comparable population between strong and weak binding sites, the photodesorption efficiency profile as a function of wavelength (Fig. 6b) is predominantly determined by the multiplication of photoabsorption cross section and quantum yield. Since these two terms cannot be separated experimentally, we can only assume that if the quantum yield is binding site independent, the wavelength-dependent efficiency trend can be regarded as the absorption spectra. To evaluate the photoabsorption cross section independently is not possible in our experiments and is out of the scope of this study.

## Acknowledgments

This work was supported by JSPS KAKENHI, grant numbers JP22H00159 and JP23H03982, and was partly supported by the Grant for Joint Research Program of the Institute of Low Temperature Science, Hokkaido University (22K002). The authors would like to thank Professor W.M.C. Sameera for sharing the quantum chemical calculation results and the fruitful discussion.